\documentclass[final,5p,times,twocolumn,authoryear]{elsarticle}

\usepackage{amssymb}
\usepackage{mathtools}

\usepackage{arydshln,leftidx}
\usepackage{amsmath}
\usepackage{amssymb}
\usepackage{kbordermatrix}
\usepackage{graphics}
\usepackage{graphicx}
\usepackage{enumitem}
\usepackage{textcomp}
\usepackage{color}
\usepackage{outlines}
\usepackage{float}

\usepackage{listings}
\definecolor{dkgreen}{rgb}{0,0.6,0}
\definecolor{gray}{rgb}{0.5,0.5,0.5}
\definecolor{mauve}{rgb}{0.58,0,0.82}
\usepackage{amsthm}

\newtheorem{definition}{Definition}[section]
\newtheorem{construction}{Construction}[section]
\newtheorem{remark}{Remark}[section]
\newtheorem{example}{Example}[section]

\newtheorem{algorithm}{Algorithm}[section]

\usepackage{times}
\usepackage{textcomp}
\usepackage{setspace}
\usepackage{bibspacing}
\usepackage{threeparttable}

\usepackage{changepage}   

\usepackage{etoolbox}
\let\bbordermatrix\bordermatrix
\patchcmd{\bbordermatrix}{8.75}{4.75}{}{}
\patchcmd{\bbordermatrix}{\left(}{\left|}{}{}
\patchcmd{\bbordermatrix}{\right)}{\right|}{}{}

\usepackage{blkarray}
\usepackage{multirow}
\usepackage{wrapfig}
\usepackage{tikz}

\usepackage{subfig}
\usetikzlibrary{matrix,fit}
\usepackage[OT1]{fontenc}

\usepackage[authoryear]{natbib}

\usepackage{wrapfig}

\makeatletter
\renewcommand\@biblabel[1]{\textbullet}
\makeatother

\journal{Ad Hoc Networks}

\begin{document}

\begin{frontmatter}

\title{A Secure and Robust Scheme for Sharing Confidential Information in IoT Systems}

\author{Lake Bu, Mihailo Isakov, Michel A. Kinsy}

\address{Adaptive and Secure Computing Systems Laboratory, Department of Electrical and Computer Engineering, \\ Boston University, Boston, USA \vspace{-.2in}}

\begin{abstract}
In Internet of Things (IoT) systems with security demands, there is often a need to distribute sensitive information (such as encryption keys, digital signatures, or login credentials, etc.) among the devices, so that it can be retrieved for confidential purposes at a later moment. However, this information cannot be entrusted to any one device, since the failure of that device or an attack on it will jeopardize the security of the entire network. Even if the information is divided among devices, there is still the danger that an attacker can compromise a group of devices and expose the sensitive information. In this work, we design and implement a secure and robust scheme to enable the distribution of sensitive information in IoT networks. The proposed approach has two important properties: (1) it uses Threshold Secret Sharing (TSS) to split the information into pieces distributed among all devices in the system - and so the information can only be retrieved collaboratively by groups of devices; and (2) it ensures the privacy and integrity of the information, even when attackers hijack a large number of devices and use them in concert - specifically, all the compromised devices can be identified, the confidentiality of information is kept, and authenticity of the secret can be guaranteed. 
\end{abstract}


\end{frontmatter}

\bigskip
\section{Introduction}
\label{sec:Introduction}
Internet of Things (IoT) and connected devices have transformed our lives. IoT systems are actively 
 deployed in a variety of settings, including homes, hospitals, battlefields, schools, airports, manufacturing plants, and more. The architecture of these systems, generally, consists of devices connected to one another or users/clients where the main network activity is data or information exchanges. In contrast to general and non-sensitive information exchange, such as reading sensors or controlling air conditioners remotely, there are many instances where critical or confidential information needs to be shared or routed among the devices. These pieces of information or ``secrets" are used by the devices or the users/clients to perform security or privacy related functions in the IoT system. The information could be encryption keys, digital signatures, login credentials, or important account numbers.
 
However, such a secret cannot be entrusted to any individual device, because the malfunction of a single device might then jeopardize the security of the entire network. Therefore, an appropriate approach is to split the secret and distribute it among multiple devices. The most commonly adopted technique in this area is threshold secret sharing (TSS). In an IoT or distributed system, TSS is generally carried out by a dealer (usually the server or administrator of the IoT system), which divides the secret and parcels those pieces among multiple holders (the devices), in such a way that the secret can only be reconstructed collaboratively by subsets of holders whose size has to reach a minimum number. This minimum size is called ``threshold." Below the threshold, the secret is theoretically safe and kept private from retrieval. 

Practical secret sharing techniques are deployed in many real world applications hat include IoT systems. The most common example is key management in wireless sensor networks. Rather than entrusting the cryptographic key to a single node, which can be easily compromised in hostile environments, the key is shared among a group of nodes and can only be retrieved collaboratively [\cite{AC2005}] to be used for digital signature or other cryptographic purposes at an other terminal. If some nodes are found to be malfunctioning, then their access will be revoked, and they will be replaced by the same number of healthy nodes to reach the threshold. One such application is the ``Vanish" project [\cite{RG2009}], which uses the threshold property to make the secret key in a distributed system vanish when the number of shareholding nodes gradually decreases to below the threshold. Another application is in Hardware Security Module (HSM) based systems. HSMs are widely used in bank card payment systems. Some HSMs [\cite{TH2013}] are produced and distributed by certification authorities (CAs) and registration authorities (RAs) to generate and share important secret keys under Public Key Infrastructure (PKI). These HSMs also require implementation of a multi-part user authentication scheme, namely threshold secret sharing. The most well-known application is probably DNS Security (DNSSEC) [\cite{DNS2010}], which ensures the DNS (Domain Name System) servers can connect users and their Internet destinations (URLs and IPs) in a secure and verified manner. Its root key is split and shared among seven holders all over the world. In the case of an attack, if any five or more of the holders are able to come to a U.S. base, then they can reconstruct the root key using their shares to restore the Internet connections. Technology survey companies also use TSS to store sensitive survey data to prevent them from being extracted by any single data analyst without the participation of others [\cite{LA2016}].    

However, although this technique reduces the risk of losing all the confidential information due to a malfunction of one of a few devices, there is still a danger when attackers compromise a larger group of them. Due to their distributed nature, TSS schemes are susceptible to a number of attacks, like passive attacks, man-in-the-middle (MITM) or share manipulations, i.e., cheating. These attacks, resulting in share disclosure or distortions, may lead to the leakage of the original secret or retrieval of a false secret. Generally speaking, the TSS is able to maintain the privacy of the secret information under the existence of a small number (below the threshold) of cheaters. However, it alone cannot guarantee the integrity of the secret. 

Although, there have been many secure TSS schemes, they are often limited in their adversarial capabilities, i.e., cheater tolerance. For instance, [\cite{CR2008}] proposed a secure version of TSS based on secret validation, which is able to detect, but not identify, any number of cheaters. The authors in [\cite{WM2016}] on the other hand, leveraged the superimposed codes with secret verification and were able to locate no more than $n^{0.5}$ cheaters, where $n$ is the total number of devices involved. With higher computation complexity, [\cite{MS1981, RG2001, FG2006}] verify the shares with error control coding (particularly Maximum Distance Separable codes) to boost the cheater tolerance to nearly $n/3$. However, when the number of cheaters exceeds their fault tolerance, neither the privacy nor the integrity of the secret can be guaranteed. In addition, the dishonest parties can even frame the honest ones as cheaters. 

Therefore, we propose a secure and robust scheme to enable the sharing of the confidential information in IoT systems with a stronger cheater tolerance. The major contributions of this work are: 
\begin{enumerate}
	\item	The proposed approach uses Threshold Secret Sharing (TSS) to split the secret into shares distributed among the devices in the system, so the secret can only be retrieved collaboratively by groups of devices; 
	\item	It adds additional security features on top of the original TSS functionality; specifically, it protects the confidentiality of the secret even when attackers have hijacked a group of devices;	
	\item It also ensures the integrity of the secret even when attackers hijack a large number of devices, collude, or manipulate the shares to forge fake secrets; 
	\item The proposed approach is able to detect and identify cheaters or compromised share holders up to a given theoretical upper bound;
	\item It provides an automation tool to aid in the secret sharing procedure and system programming based on user-specified parameters. 
\end{enumerate}

For evaluating the feasibility of the proposed robust secret sharing approach in systems consisting of physically distributed and connected devices, we introduce the Odysseus IoT open-interface testbed system. Testing on the Odysseus IoT testbed serves to validate the practicality of the attack models and associated defenses. It also highlights how a practical secure information sharing mechanism may be implemented. 

Section II introduces the details of the Odysseus IoT testbed system, as well as the original threshold secret sharing scheme. The section also covers the attack model. Section III summarizes some existing secure protocols for the TSS. Section IV follows up with the vulnerabilities of those protocols under the attack model. Section V describes the proposed secure and robust secret sharing scheme, as well as a cheater identification protocol. Section VI presents the design automation tool, and finally, Section VII concludes the paper.

\section{The Odysseus IoT System, the Original Threshold Secret Sharing Scheme, and the Attack Models}
\label{sec:Related-Works}

In this section, we first introduce the Internet of Things (IoT) Testbed System, ``Odysseus", on which we evaluate the practicality of the proposed secure TSS scheme. We use this system to introduce and illustrate the proposed approach without a loss of generality, and to provide some deployment concreteness. 

\subsection{System Model - Odysseus IoT}
The original motivation of developing a secure and robust TSS is to protect systems like the Odysseus IoT system. In such a system, the dealer is the service provider, which provides the Odysseus boards and is responsible for their deployment. The Odysseus boards are sensor hosting boards supporting various types of sensors. The boards have wireless communication modules for data exchange. The clients or users can pick the sensors to be installed and processed on the boards via GPIO ports before the boards' deployment. These sensors can either be heterogeneous or homogeneous. One example of Odysseus' application is in fire-fighting and rescue: heat sensors to map the fire intensity and location within a burning building, and motion sensors to identify human presence. 

In general, the dealer (administrator) of the Odysseus system can deploy a large number of sensor boards to an area, and their sensor data can be requested remotely by different clients. From time to time, a client will request sensing data from a group of sensors, while retrieving from them a secret, if necessary. The secret, such as an encryption key, will be used by the client on various applications associated with the sensor data. The system chart and prototype of Odysseus are shown below in Fig. \ref{fig:Odysseus} and \ref{fig:prototype}. 
\begin{figure}[h] 
	\begin{center} 
		\includegraphics[width=3.5in]{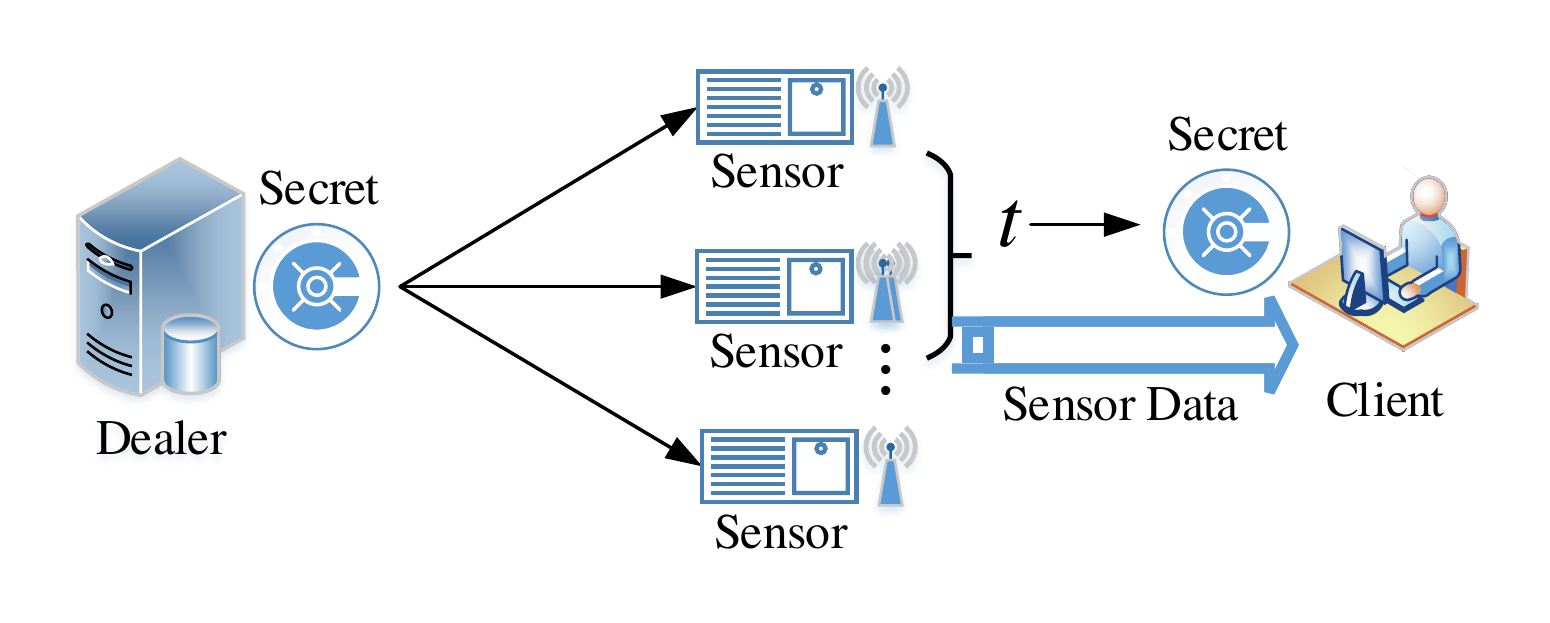} 
		\caption{\small The three layers of the Odysseus system: the dealer who deploys the boards and the secret, the sensor boards as the shareholders with wireless communication capability, and the client(s) who collects the data as well as the secret.  
			\label{fig:Odysseus}} 
		\vspace{-0.1in}
	\end{center} 
\end{figure} 

\begin{figure}[h] 
	\begin{center} 
		\includegraphics[width=3.0in]{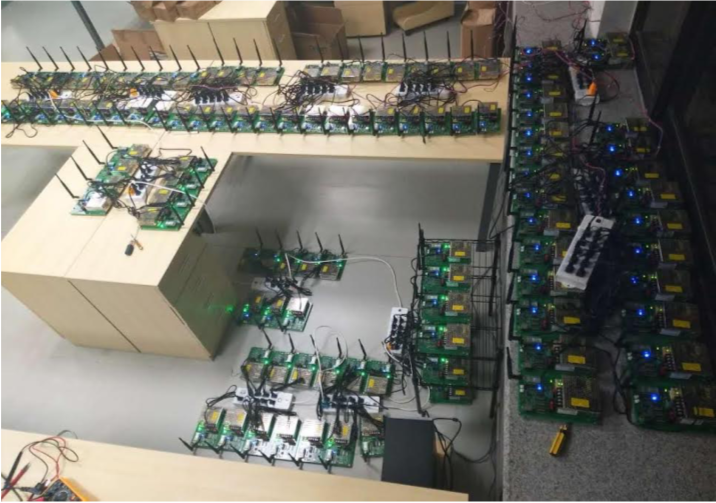} 
		\caption{\small The prototype boards of Odysseus. 
			\label{fig:prototype}} 
		\vspace{-0.2in}
	\end{center} 
\end{figure} 

The security of this IoT system also needs to be addressed. Although the dealer and clients can be trusted, the sensor hosting boards scattered all over a region are not physically monitored. Since any number of them can be subject to passive or active attacks, no critical information such as the secret can be entrusted to any individual board. There is even a danger of a large number of them being hijacked by attackers, who might thereby gain full access to those devices. Therefore, a secure protocol to maintain the privacy and integrity of the secret is needed, as well as error tolerance to deal with the existence of compromised boards. 

\subsection{The Original Threshold Secret Sharing}
\label{sec:Original}
As noted in Section 1, TSS divides confidential information between devices, instead of storing the whole secret on each device, such that a defect in, or the compromise of, a single device will not impair the security of the entire network.

The following notations are used to describe and evaluate the original threshold secret sharing scheme, as well as the related secure variations:
\vspace{-0.1in}
\addtolength{\leftskip}{0mm}
\begin{itemize}[leftmargin=\dimexpr\parindent+1mm+0.5\labelwidth\relax]
	\renewcommand{\labelitemi}{\scriptsize$\bullet$} 
	\setlength\itemsep{0.05em}
	\item $ S $: the original secret (a piece of confidential information);
	\item $ D_{i} $: the public ID of the $i^{th}$ shareholder;
	\item $ h_{i} $: the share of $S$ of the $i^{th}$ shareholder;
	\item $ t $: the threshold of a secret sharing scheme;
	\item $ c_{est} $: the number of estimated cheaters;
	\item $ c_{act} $: the number of actual cheaters;
	\item $ n $: the total number of shareholders involved in a computation;
	\item $ b $: the number of bits in a vector variable;
	\item $\oplus$:  the addition operator in finite fields;
	\item $\cdot$ :  the multiplication operator in finite fields;
	\item $\bigoplus$: the cumulative sum operator in finite fields;
	\item $\prod$:	 the cumulative product operator in finite fields;
	\item $\mathtt{\sim}$ : the distortion of a vector;
	\item $ MAC() $: a secure message authenticating function;
	\item $ ENC() $: a cryptographic encryption function; 
	\item $ EtM() $: an Encrypt-then-MAC function; 
	\item $ K $: the cryptographic key;
	\item $||$: the concatenation operator;
	\item $ E $: the encoded secret where $E = EtM(S, K)$;
	\item $ P_{miss} $:  the probability of failing to detect cheating in the IoT system.
\end{itemize}

The concept of $t$-threshold secret sharing (TSS) was first introduced by Shamir [\cite{AS1979}]. He argued that all computations should be carried out over Galois finite field ($GF$) arithmetic, in order to maintain the information's theoretical security. To share a secret $S$, a polynomial of degree $(t-1)$ is used to compute and distribute the shares, where the secret $S$ serves as the free or leading coefficient, and all other coefficients can be arbitrarily chosen. The shares are the evaluations of the polynomial by each holder's  $D_i$. 

The share distribution equations when $S$ is placed as the free coefficients is: 
\begin{equation*}  
h_{i} = S \oplus a_{1}D_{i} \oplus a_{2}D_{i}^2 \oplus \cdots \oplus a_{t-1}D_{i}^{t-1}.
\end{equation*}
And as the leading coefficient: 
\begin{equation} \label{eq:1}
h_{i} = a_{0} \oplus a_{1}D_{i} \oplus a_{2}D_{i}^2 \oplus \cdots  \oplus SD_{i}^{t-1} .
\end{equation}
where $S$, $h_i$, $D_i$ $\in GF(2^b)$. 

The ID numbers, $D_i$, are publicly known, while the share, $h_i$, are kept private by shareholders. 

With any subset of at least $t$ shareholders' IDs and shares, one can use the Lagrange interpolation formula to reconstruct the secret. 

If $S$ is placed at the free coefficient, it can be retrieved by:
\begin{equation*} 
S = \bigoplus_{i=0}^{t-1} {\frac{D_{i} \cdot h_{i}}{\prod_{j=0,j \neq i}^{t-1} {(D_{i} \oplus D_{j})}}}.
\end{equation*}

If $S$ is the leading coefficient, it can be retrieved by: 
\begin{equation} \label{eq:2}
S = \bigoplus_{i=0}^{t-1} {\frac{h_{i}}{\prod_{j=0,j \neq i}^{t-1} {(D_{i} \oplus D_{j})}}}.
\end{equation}

Such a construction is $(t-1)$-private. This means it needs at least $t$ shareholders to reconstruct the secret and so any $(t-1)$ or fewer shareholders have zero knowledge of the secret.

For computational simplicity, in this paper we choose to place $S$ as the leading coefficient as shown in [Eq. \ref{eq:1} and \ref{eq:2}]. We also assume that the system works over finite field $GF(2^b)$, where, as in most computer systems, $b = 32, 64, 128, 256, \cdots$.

\begin{figure}[h] 
\begin{center} 
\includegraphics[width=3.5in]{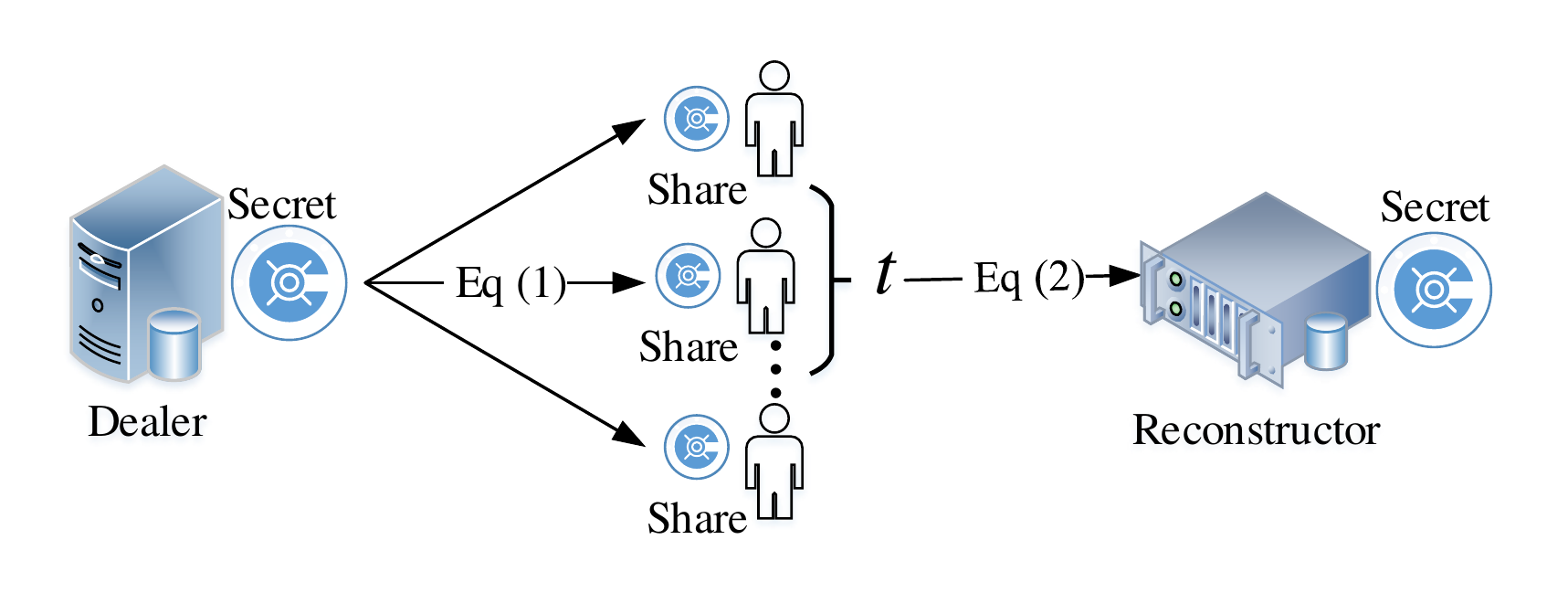} 
\caption{\small The secret sharing and reconstruction flow. The reconstructor can be omitted if there is no end user and every shareholder (either device or person) has trustworthy computation capability. \label{fig:original}} 
\vspace{-.2in}
\end{center} 
\end{figure}

The original scheme's share distribution and secret reconstruction procedures are shown in Fig. \ref{fig:original}, which matches with the Odysseus and many other IoT architectures very well in the administrator - devices - clients three layer structure.  

\begin{remark}
\normalfont Shamir's secret sharing scheme is supposed to work under finite field arithmetic where the field size should be a prime or a power of a prime. Ordinary arithmetic will be vulnerable and any secret can be retrieved by at most two carefully selected shareholders instead of $t$. 

In ordinary positive integer arithmetic, for instance, if a shareholder's ID is $D_{i} = 1$, this holder's share will be $h_{i} = a_0 + a_1 + \cdots + S$, namely the sum of the all coefficients of (\ref{eq:1}). And in ordinary arithmetic, it is obvious that $h_{i} > a_{l} | a_{l} \in \{a_0, a_1, \cdots, S\}$. Then, this holder can find another holder with ID $D_{j} \geq h_{i}$ whose share is $h{j}$. If these two shareholders work together, they can easily uncover the secret, regardless of $t$, by expressing $h_{j}$ in the radix of $D_{j}$, where the most significant digit will be $S$.

However, in finite field or modular arithmetic, one can never have $h_{i} > a_{l} | a_{l} \in \{a_0, a_1, \cdots, S\}$ if $h_{i} = a_0 \oplus a_1 \oplus \dots \oplus S$.
 \hfill\(\square\)
\end{remark}

\subsection{Attack Model} 
\label{sec:attack-models}
We define an attack model below, which is much stronger than what the original scheme and its conventional secure variations can handle.

\begin{definition} \label{def:attack-model}
	\normalfont The attack model in this paper is described by the following characteristics:
	\begin{enumerate}
		\item The dealer and the clients are trusted;
		\item The shareholders (devices in an IoT system) are not trusted and there is no limit to the number of compromised devices or cheaters. 
		\item The cheaters are able to gain full control of hijacked devices, meaning cheaters can read memory contents, use IO ports, or tamper with devices.
		\item The cheaters can also eavesdrop or tamper with the communication channels between the devices, the dealer and/or the clients.
		\item The attackers have the knowledge of the system's basic parameters ($n, t$, equations [\ref{eq:1}, \ref{eq:2}] etc.). They can work collaboratively.
		\item The goals of the attackers are: 
		\begin{enumerate}
			\item \textit{Passive attack}: to compute and acquire the original secret stealthily;
			\item \textit{Active attack}: to select their own secret and submit it to the clients without being spotted.
		\end{enumerate}
	\end{enumerate}
\end{definition}

\textbf{\textit{Note:}} Besides the shares, each Odysseus board also submits its sensor data to the clients. However, because those are source data, their verification is another issue beyond the scope of this paper. 

When the cheaters work collectively, they are able to share any information they hold, or to modify it according to their common interest. We also assume that the cheaters have sufficient computational power to calculate equations such as [\ref{eq:1}, \ref{eq:2}] and other necessary tasks.

\bigskip
\vspace{-.2in}
\section{The Conventional Secure Protocols for TSS}
\label{sec:current}
In this section, we present some of the existing secure protocols and their associated passive and active attack models. We also highlight their vulnerabilities under our attack model. 

\subsection{Against Passive Attacks}
The key property of TSS is that it only allows $t$ or more shareholders (devices) to retrieve the secret. Below this threshold, the secret information is theoretically secure. Namely, $t-1$ devices have no more knowledge of the secret than any individual device does. However, if the cheaters have compromised $t$ or more devices, so $c_{act} \geq t $, then the privacy of the secret is not guaranteed, since they can use [Eq. \ref{eq:2}] to retrieve it. 

\subsection{Against Active Attacks}
\label{sec:convention}
Soon after the introduction of the Shamir's secret sharing scheme, it was noticed that if any number of the shareholders participating in the secret reconstruction apply an active attack by changing their shares to make $h_i$ to $\tilde{h_i} \neq h_i$, the retrieved secret will be distorted $\tilde{S} \neq S$ according to [Eq. \ref{eq:2}]. Therefore the authenticity of the submitted shares or the retrieved secret needs to be verified. 

\subsubsection{Share Verification}
\label{sec:secret-verify}
Researchers [\cite{MS1981, RG2001, FG2006}] have proposed approaches to verify the validity of shares with a probability of 1. The common feature in these approaches is that if the shares can be encoded to a specified error control code (ECC) codeword. The codeword's symbols, i.e., shares can be verified and corrected up to the ECC's capability. 

Particularly, the share distribution [Eq. \ref{eq:1}] is inherently equivalent to the non-systematic encoding equation of the well-known Reed-Solomon (RS) ECC codes. RS codes are maximum distance separable (MDS) codes which meet the Singleton bound with equality. With such a distribution equation, an ($n, t, d$) Reed-Solomon codeword ($h_{0}, h_{1}, \cdots h_{n-1}$) is encoded with $n$ symbols (shares) in total, $t$ information symbols, and distance $d = n-t+1$, which can correct up to $\frac{d-1}{2}$ (or $\frac{n-t}{2}$) erroneous symbols with the algorithms in [\cite{EB2015, SG2003}].

In the secret sharing language, with $n$ shareholders' IDs and shares, we are able to tolerate up to $c_{est} \leq \frac{n-t}{2}$ shares maliciously modified by cheaters. Theoretically speaking, the error correction capability of RS codes can tolerate up to $c_{est} < n/2$ cheaters if $n \gg t$. However, commonly, the assumption that there should be $c_{est} < t$ cheaters is made, such that a group of all cheaters have no access to the secret [\cite{HK1993}]. Then we have
\begin{equation} \label{eq:3}
c_{est} < n/3.
\end{equation}

If $n$ instead of $t$ shareholders are involved in the share error correction by RS decoders, then the correctness of the retrieved secret is ensured when [Eq. \ref{eq:3}] holds. Consequently, the secure secret sharing is both $(t-1)$-private and $(t-1)$-resilient; that is, up to $t-1$ shareholders cannot reconstruct the secret, and up to $t-1$ cheaters cannot affect the correctness of the secret [\cite{LM2015}].  

\subsubsection{Secret Verification}
Besides share verification with a share correction probability of 1, another approach is to sign the original secret with a key $K$ using a message authentication code (MAC) function. Then the original secret is shared together with its MAC (usually in a manner of concatenation) to the holders. Denoting the encoded secret as $(S || MAC(K, S))$, then [Eq. \ref{eq:1}] becomes:
\begin{equation} \label{eq:4}
h_{i} = a_{0} \oplus a_{1}D_{i} \oplus a_{2}D_{i}^2 \oplus \cdots \oplus (S || MAC(K, S))D_{i}^{t-1} .
\end{equation}

At the reconstructor end, after the retrieval of the possibly distorted $(\tilde{S} || \widetilde{MAC(K, S)})$, the following authentication equation is evaluated: 
\begin{equation} \label{eq:5}
MAC(\tilde{K},\tilde{S}) \stackrel{?}{=} \widetilde{MAC(K, S)}.
\end{equation}

An inequality detects cheating. If this MAC function has a high enough security level, such as a collision or mis-detection probability of $2^{-128}$ (or lower), then it is generally believed that all distortions will be spotted. The secure protocol of the Shamir's secret sharing is shown below.
\begin{figure}[h] 
	\begin{center} 
		\includegraphics[width=3.5in]{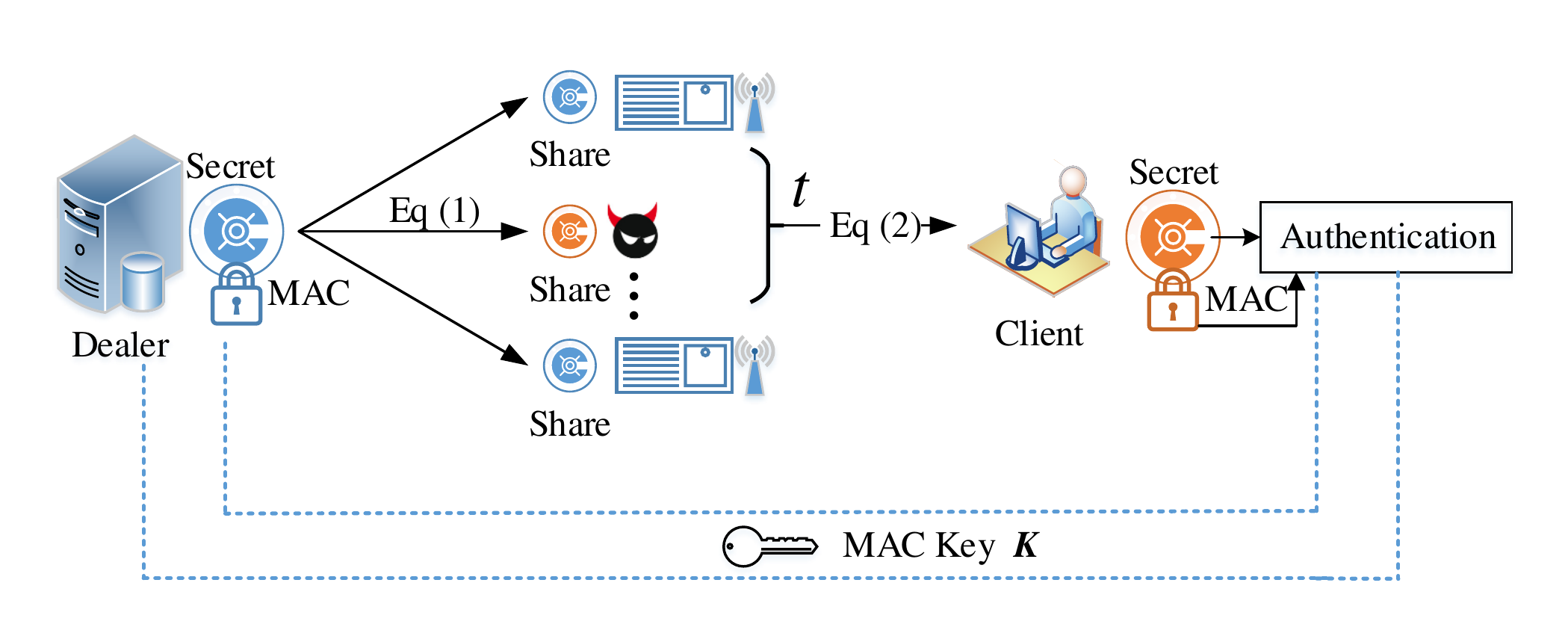} 
		\caption{\small The secret sharing scheme with secret authentication in the context of Odysseus system. 
			\label{fig:signed}} 
		\vspace{-0.2in}
	\end{center} 
\end{figure} 

There are two common approaches for signing the original secret: HMAC with a key and AMD codes with a random vector. 

\paragraph{A. HMAC with a Key} \hfill \break
HMAC, keyed-hashing for message authentication code, is the most often used technique for authentication today. To sign a secret $S$, the nested equation is defined as follows [\cite{RFC2104}]:
\begin{definition}
	\normalfont Let $HMAC()$ be the HMAC function, $K$ the signing key, and $K'$ be derived from $K$ by padding to the right zeros to the block size. Also let $H$ be a hashing function, $opad$ the outer padding and $ipad$ the inner padding. Then:
	\begin{equation} \label{eq:6}
	HMAC (K, S) = H \big( (K' \oplus opad) || H((K' \oplus ipad) || S) \big)
	\end{equation}
\end{definition}

The client can authenticate the secret using the HMAC version of [Eq. \ref{eq:5}]:
\vspace{-0.1in}
\begin{equation} \label{eq:7}
HMAC(\tilde{K}, \tilde{S}) \stackrel{?}{=} \widetilde{HMAC(K,S)}.
\end{equation}

With SHA-2 256 or higher used for $H()$ [\cite{RFC6234}], the collision rate is less than $2^{-128}$, and thus, it is considered cryptographically secure.  

\paragraph{B. AMD with a Random Number} \hfill \break
[\cite{CR2008}] have proposed an Algebraic Manipulation Detection (AMD) code to detect any modification of secrets with a probability close to 1. [\cite{WK2011}] later generalized this code with a flexible construction. 

Unlike HMAC, it operates over finite fields and its security level is adjustable by block size $b$. The AMD encoding is defined as follows

\begin{definition}
	\normalfont Let $K = (K_1, K_2, \cdots , K_m)$, where $K_i \in GF(2^b)$ is a randomly generated $b$-bit vector. An $g^{th}$ order Generalized Reed-Muller code ($GRM$) with $m$ variables consists of all codewords ($f(0), f(1), \cdots , f(2^{bm}-1)$), where $f(K)$ is a polynomial of $K = (K_1, K_2, \cdots , K_m)$ of degree up to $g$. 
	 
	Let
	\vspace{-.05in}
	\begin{equation*}
	A(K)=  
	\begin{cases} \bigoplus_{i=1}^{m} K_i^{g+2}, & \text{if $g$ is odd;}
	\\
	\bigoplus_{i=2}^{m-1} K_1K_i^{g+1}, & \text{if $g$ is even and $m>1$;}
	\end{cases}
	\end{equation*}
	where $\bigoplus$ is the accumulated sum in $GF(2^b)$. Let
	\vspace{-.05in}
	\begin{equation*}
	B(K, S)= \bigoplus_{1 \leq j_1 + j_2 + \cdots + j_1 \leq g+1} y_{j_1, j_2, \cdots, j_m} \prod_{i=1}^m K_i^{j_i},
	\end{equation*}
	where $\prod_{i=1}^m K_i^{j_i}$ is a monomial of $R$ of a degree between 1 and $g+1$. And $\prod_{i=1}^m K_i^{j_i} \notin  \triangle B(K, S)$ which is defined by:
	\begin{equation*}
	\begin{cases} \{K_1^{h+1}, K_2^{g+1}, \cdots, K_m^{g+1}\}, \text{if $g$ is odd;}
	\\
	\{K_2^{g+1}, K_1K_2^{g}, \cdots, K_1K_m^{g}\}, \text{if $g$ is even and $m>1$}.
	\end{cases}
	\end{equation*}

	Let $f(K, S) = A(K) \oplus B(K, S)$, then a generalized AMD codeword is composed of the vectors $(S, K, f(K, S)) $, where $S$ is the information portion, $K$ the random vector, and $ f(K, S) $ the redundancy signature portion [\cite{WK2011}]. \hfill\(\blacksquare\) 
\end{definition}

\begin{remark}
	\normalfont If the attack involves a non-zero error on the information $S$, which is the major purpose of almost all attacks, then in $f(K, S)$ the term $A(K)$ can be omitted [\cite{BK2017}]. Furthermore, if only one random number vector is used, the encoding equation can be simplified to
	\begin{equation} \label{eq:AMD}
	AMD(K, S) = f(K, S ) = \bigoplus_{1 \leq j_1 + \cdots + j_i + \cdots + j_m \leq h+1} S_{j_1, \cdots, j_i, \cdots, j_m} K^{j_i}
	\end{equation}
	where $S_{j_i}$ is a $b$-bit block of $S$.  \hfill\(\blacksquare\)
\end{remark}

The client can authenticate the secret using the AMD version of [Eq. \ref{eq:5}]:
\begin{equation} \label{eq:9}
AMD(\tilde{K}, \tilde{S}) \stackrel{?}{=} \widetilde{AMD(K,S)}.
\end{equation}

The probability of mis-detecting a distortion of $S$ in [Eq. \ref{eq:9}] has an upper bound, $\frac{g}{2^b}$ [\cite{CR2008}], where $g$ is a very small number in most constructions. With 128 bits (or larger) selected as $b$, the security level of AMD codes will be on the same order of HMAC ($2^{-128}$ or less in attack mis-detection rate).

\textit{\textbf{Note:}} Although HMAC and AMD codes are different approaches for authenticating the retrieved secrets, there is no essential difference in their design philosophy as [Eq. \ref{eq:7}] and [Eq. \ref{eq:9}] have shown. 

It should be noted that there are two potential drawbacks to the secret verification approach. First, no method for transmitting the MAC key, $K$, from the dealer to the client is specified for either approach. In addition, while these approaches can detect the distortion of the secret, they cannot identify the cheaters or restore the correct secret.

\bigskip
\section{Vulnerabilities of the Conventional Secure TSS Schemes}
\label{sec:vulnerable}
In this section, we illustrate the vulnerabilities associated with conventional secure schemes under the attack model defined previously. Because of the distributed nature of IoT systems, it is not unusual to have attacks of a scale unanticipated by designers. The demand for more secure and robust confidential information sharing scheme for IoT systems is the main motivation for the approach proposed in the next section.

\subsection{Passive Attack: Acquiring the Original Secret}
Usually an assumption has to be made that $c_{est} < t$ so that a group of all cheaters cannot retrieve the secret by themselves. However, a case with more estimated cheaters such that $c_{act} \geq t > c_{est}$, could exist. With any $t$ of them, it is easy to acquire the original secret by [Eq. \ref{eq:2}].

\subsection{Active Attack: Making the Secret Unaccessible}
Here we assume the IoT system's TSS is already equipped with a share verification module. As mentioned in Section 3.2.1, the essence of such module is to encode the shares into a codeword, whose validity can be verified by the RS decoding algorithm. Although RS codes are known for their strong error correction (tolerating $c_{est} < n/3$ cheaters), their encoding procedure is linear and thus, susceptible to cheating exploits. 

If the number of cheaters satisfy $(n/3 < c_{act} < n-t+1)$, although the RS decoder can still raise an alarm for cheating, it is already beyond the share error correction capability of the RS code. Therefore the system is unable to retrieve the secret or identify the cheaters.

\subsection{Active Attack: Forging a Legal Secret}
\label{sec:forge}
If the number of cheaters satisfies $ (n-t+1 \leq c_{act} \leq n) $, they will be able to manipulate the entire system. For instance the cheaters can pick a share distribution polynomial different from [Eq. \ref{eq:1}] with random coefficients $b_i$ and their own forged secret $\tilde{S}$: 
\begin{equation} \label{eq:10}
h'_{i} = b_{0} \oplus b_{1}D_{i} \oplus b_{2}D_{i}^2 \oplus \cdots \oplus \tilde{S}D_i^{t-1} 
\end{equation}

The new shares $h'_{i}$ of the cheaters will be the evaluation of [Eq. \ref{eq:10}] by the same IDs $D_i$. When $c_{act} \geq n-t+1$, the cheaters' shares will form a new legal RS codeword which will never be detected by the RS decoder. The secret reconstruction will then submit the cheaters' secret $\tilde{S}$ to the client. If the client uses it on his/her own important applications such as digital signatures, the attackers can effortlessly break those applications. 

\begin{example}
	\label{example1}
	\normalfont A secret sharing system has a secret $S = 111$ in the $GF(2^3)$ finite field. It requires $t=2$ shareholders to reconstruct the secret every time. The following share distribution polynomial is used to generate the shares:
	\[
		h_{i} = a_{0} \oplus SD_{i} = 010 \oplus 111D_{i}.
	\]
	
	The protocol is designed in such a way that up to 1 cheater can be tolerated. Therefore, in the secret reconstruction stage there will be $n = 3c_{est}+1 = 4$ shareholders involved. Suppose that in the secret reconstruction, shareholders with IDs $ D_0 = 001, D_1 = 010, D_2 = 011, D_3 = 100 $ are involved. And the shares distributed to them are $ h_0 = 101, h_1 = 111, h_2 = 010, h3 = 001$. These 4 shares form a legal RS codeword $v = (101, 111, 010, 001)$ with distance $d=n-t+1=3$ and it can correct up to 1 error.  
	
	Now all 4 of them are cheating collusively, and they have selected their own secret $\tilde{S} = 100$ and a different share distribution polynomial:  
	\[
		h'_{i} = b_{0} \oplus \tilde{S}D_{i} = 001 \oplus 100D_{i}.
	\]
	
	Thus their shares will be maliciously changed to $h_0 = 101, h_1 = 010, h_2 = 110, h3 = 111$, which is also a legal codeword $v' = (101, 010, 110, 111)$ of a $(n, t, d) = (4, 2, 3)$ RS code. This codeword will unfortunately be considered as a valid codeword by the RS decoding algorithm [\cite{SG2003}] and there will be no cheating alarm. As a result, the fake secret $\tilde{S} = 100$ is retrieved by those shares under [Eq. \ref{eq:2}]. During the entire procedure the cheating will not be detected.  \hfill\(\square\)
\end{example}

\subsection{Active Attack: Framing Honest Shareholders}
\label{sec:framing}
Another vulnerability that cheaters can exploit when $ (n-t+1 \leq c_{act} \leq n) $ is to frame honest shareholders, so that the decoder treats the honest parties as ``cheaters" and cheaters as ``honest shareholders." 
If $c_{act}$ is large enough that the number of honest shareholders is $ n-c_{act} \leq \frac{n-t}{2} $, then the honest shareholders are within the RS decoder's error correction capability. Since all of the cheaters' shares are generated by the same forged secret sharing polynomial, the honest minority will be treated as cheaters and ``corrected." The cheaters' fake secret will be regarded as the valid secret as the result of [Eq. \ref{eq:2}]. 

\begin{example}
	\normalfont Suppose that we have the same secret sharing system as in Example \ref{example1}. Let us have three shareholders \{$D_0 = 001, D_1 = 010, D_2 = 011$\} as cheaters, and shareholder $ D_3 = 100 $ is an honest participant. The codeword for the shares submitted to the RS decoder will be $v' = (101, 010, 110, 001)$. $v'$ will be decoded as $(101, 010, 110, 111)$ which is the cheaters' codeword. Shareholder $D_3=100$ will be labeled as a ``cheater". Consequently, the forged secret $\tilde{S}=100$ (as in Example \ref{example1}) 
	will be retrieved. \hfill\(\square\)
\end{example}

\subsection{Active Attack: Against Secret Verification}
As mentioned above, one can design an IoT with secret verification TSS capabilities. Although such a design has a high probability of detecting any number of share distortions, it alone is not able to identify the cheaters nor correct the shares. In addition, there is another problem that needs to be addressed: how to securely pass the MAC key $K$ from the dealer to the client (as in Fig. \ref{fig:signed}) in order to conduct the secret authentication, giving that the transmission channel might be eavesdropped. 

There can be more types of attacks besides the ones listed above. Especially when the number of cheaters is beyond estimation, the entire system can be subject to total manipulation. Therefore there is a demand for a more secure and resilient scheme to handle the severe attacks.

\bigskip
\vspace{-.2in}
\section{A Secure and Robust Secret Sharing Scheme for IoT}
\label{sec:Proposed}
In this section, we propose a new secure and robust secret sharing scheme for IoT systems. Compared to the current secret sharing scheme, which has limited protection against cheaters, the advantages of the proposed scheme are
\begin{enumerate}
	\item The proposed scheme protects both the confidentiality and the integrity of the secret;
	\item The proposed scheme is able to detect and identify the cheaters up to the theoretical upper bound;
	\item The proposed scheme uses the Physical Unclonable Functions (PUF) to ensure the security of the cryptographic key update;
	\item The proposed scheme works in an adaptive manner, such that a more powerful module will only be activated when the previous module fails. Thus, the scheme functions in a cost-efficient way and consumes a minimum of resources on average.  
\end{enumerate}

The following subsections are organized to present an overview of the proposed scheme, a detailed introduction of the modules of this scheme, and, finally, a simple numeric example to demonstrate the scheme.

\subsection{Overview of the Proposed Secure Secret Sharing Scheme}

The proposed scheme has four stages to ensure the basic functionality and authenticity of the secret sharing. 

\textbf{Stage 1: Dealer - Encoding and Distribution of the Secret} \\
First, the dealer will encode the secret $S$ with an Encryption-then-MAC function $EtM()$ to $E = EtM(K, S)$, where $K$ is randomly picked from the dealer's repository, which stores the challenge and response pairs (CRPs) of the client's PUF. Then the dealer distributes $E$ using [Eq.\ref{eq:2}] to $n$ shareholders. The detailed key transmission protocol will be introduced in later subsections. 

\textbf{Stage 2: Client - Secret Retrieving} \\
The client will select an arbitrary set of $t$ shareholders to participate in the secret retrieving using [Eq. \ref{eq:2}]. The retrieved secret will be authenticated by [Eq. \ref{eq:7} or \ref{eq:9}] by the $K$ generated at the client end. If the authentication claims the secret is valid, then it is considered a successful secret reconstruction with no cheating. If not, the scheme moves to Stage 3 for share correction. 

\textbf{Stage 3: Client - Share Error Correction} \\
This stage uses the Reed-Solomon error correction module in the classic protocol. Here, $n = 3c_{est} + 1$ shareholders will be invited to participate in the protocol, where $c_{est}$ is the number of estimated cheaters defined by the system. The RS decoder will try to correct the shares and then send them back to the secret reconstruction and verification modules at the client end. If the protocol passes both the share correction (by the RS decoder module) and secret verification (by the authentication module), then the secret reconstruction is successful. When $c_{act} < n/3$, the cheater tolerance probability is $100\%$. If either module fails, then the protocol ascends to its fourth stage, indicating that the actual number of cheaters is greater than $n/3$.

\textbf{Stage 4:  Client - Group Testing} \\
This stage will be activated if the previously retrieved secret is not legal. It will involve up to $n$ shareholders, among whom there are at least $n/3$ cheaters. The client will generate a group testing pattern which is able to identify up to $ c_{est} = n - t $ cheaters with a minimum number of $t$ honest holders. Even if there are more than $n - t$ cheaters, it is still able to detect the cheating, although the correct secret is beyond reconstruction because there are not enough honest holders.

The work flow of the proposed scheme is shown below.
\begin{figure}[h] 
	\begin{center} 
		\includegraphics[width=3.5in]{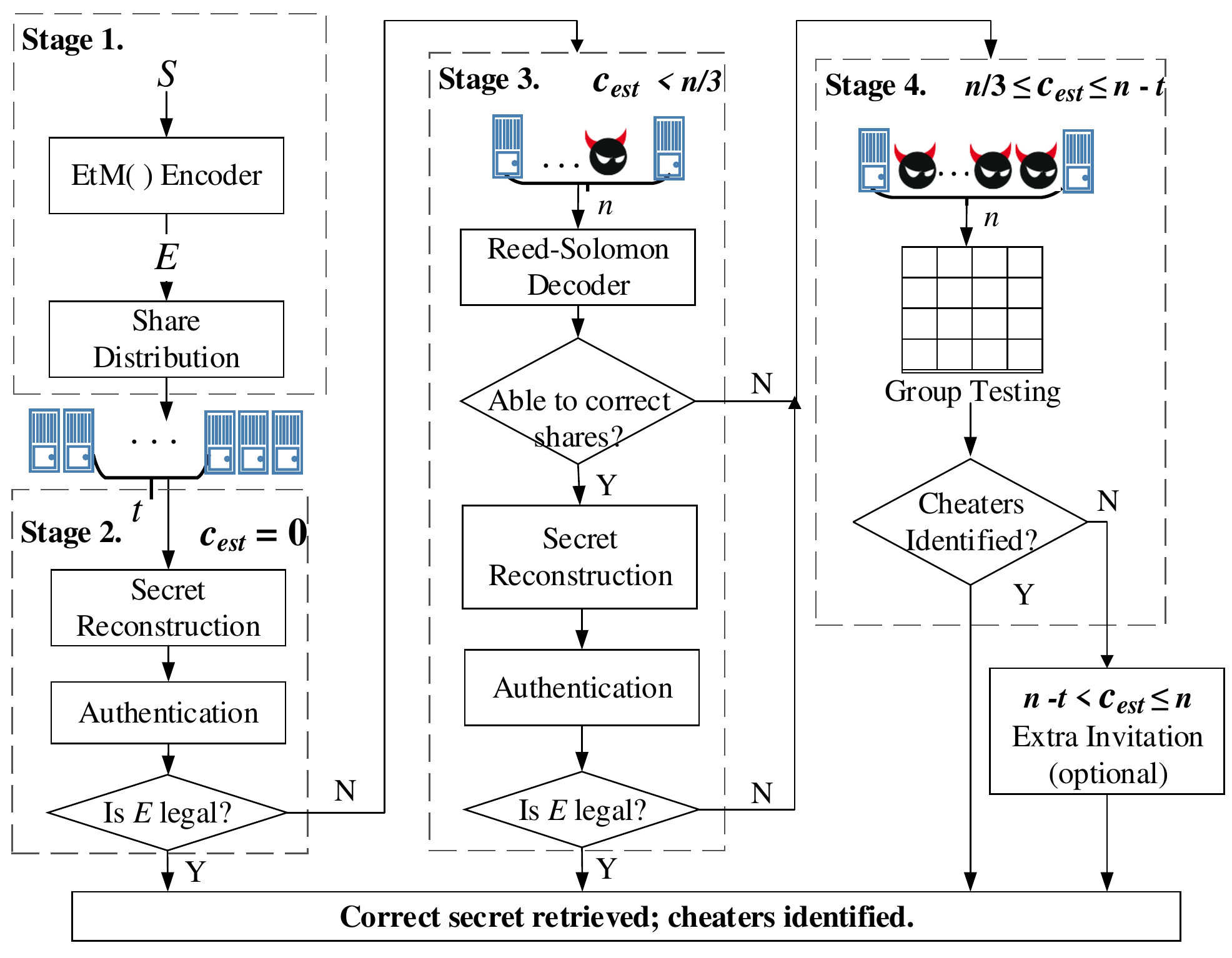} 
		\caption{\small \textbf{Stage 1} and \textbf{2} are sufficient if the number of actual cheaters $c_{act} = 0$. If cheating is detected by \textbf{Stage 2}, then \textbf{Stage 3} with RS decoder is called under the assumption of $c_{est} < n/3$. If \textbf{Stage 3} fails then \textbf{Stage 4} with group testing is able to identify $n/3 \leq c_{est} \leq n - t$ cheaters and retrieve the correct secret. If $c_{act}$ is even beyond this scale, an additional invitation module can be introduced to resolve the issue.
			\label{fig:Flowchart_PUF}} 
		\vspace{-.2in}
	\end{center} 
\end{figure}

\subsection{Secret Encoding}
In order to perform the obscuration and authentication of the secret, we will apply the Encryption-then-MAC function to encode the original secret $S$ to $E = EtM(K, S) = ENC(K, S) || MAC(K, ENC(K, S))$. The encryption function $ENC()$ can be the standard AES or other lightweight approaches. And the $MAC()$ function can be either HMAC with fixed security level $P_{miss}$, or AMD codes with flexible $P_{miss}$ as mentioned in Section \ref{sec:secret-verify}. AMD codes are able to trade off between the security level and hardware cost by adjusting the vector size $b$, which can be an ideal choice for IoT systems with limited resources. Therefore, AMD may be a better choice for this class of systems.

For some distributed systems without a client end, it is not possible to maintain the confidentiality of the secret if more than $t$ devices are compromised. This is because the TSS scheme for this case entrusts the secret to the devices themselves. However, for other IoT systems with a client end like the Odysseus, it is possible to protect the privacy of the secret even if $c_{act} \geq t$ with the help of the client. Even if the attackers have compromised more than $t$ devices, they will only acquire the cipher, but not the secret's plaintext. 

However, there is a critical issue of transmitting the EtM key to the client securely, which we discuss below. 
 
\subsection{EtM Key Transmission}
The core of this proposed scheme's security is to establish a secure transmission channel for $K$ that is
\addtolength{\leftskip}{0mm}
\begin{itemize}[leftmargin=\dimexpr\parindent+3mm+0.5\labelwidth\relax]
	\renewcommand{\labelitemi}{\scriptsize$\bullet$} 
	\item \textbf{Eavesdrop resistant}: if the cheaters eavesdrop on the channel, they should not acquire any knowledge of $K$; \vspace{-0.1in}
	\item \textbf{Easy to update}: it should be easy and secure to update $K$ on both the dealer and the client sides;  \vspace{-0.1in}
	\item \textbf{Unforgeable}: a cheater should not be able to predict, duplicate, or forge the keys; \vspace{-0.1in}
	\item \textbf{Unique}: in the case of a multi-client secret sharing system, different clients should have different sets of keys.  \vspace{-0.1in}
\end{itemize}

Based on the criteria above, a Physical Unclonable Function (PUF) based approach is an excellent and fitting solution. Another choice is to use public and private key pairs. Considering that the Odysseus and many other IoT systems are hardware based, and so it is very convenient and natural to implement PUFs on them, we will use PUFs to facilitate the transmission of $K$ in this paper. Although the concept of PUF has been known since 1983 [\cite{DB1983}], the term PUF only came into wide use in 2002 [\cite{BG2002}]. A PUF is a piece of hardware that produces unpredictable responses upon challenges due to manufacturing variations. PUFs are both easy to make and hard to duplicate, even when the exact same circuit layout and manufacturing procedure are used. A PUF can be made from a device's memory cells or circuits without modifying the device's architecture. Because of its attributes of randomness and uniqueness, PUF provides an inexpensive and integrated solution for random number or secret key generation, dynamic authentication, and identification [\cite{MY2016}].

The PUF serves as a cryptographic primitive in the manner of challenge-response pairs (CRPs). Each PUF's output (response) is a non-linear function of the outside input (challenge) and the PUF’s own physical, intrinsic, and unique diversity, its ``Silicon Fingerprints" [\cite{EE2010}]. Given the same challenge, the same PUF design on different circuits will return different responses, which cannot be predicted by just having the challenge vector. Therefore, PUF is an ideal choice in facilitating the transmission of $K$.

\subsubsection{Key Transmission Protocol}
\begin{algorithm} \label{alg: CRP}
\normalfont For the $k^{th}$ round of secret sharing, denote the secret as $S_k$, the arbitrarily selected challenge and response of the client's PUF as $CHL_k$ and $K_k$ respectively. Then the EtM key $K$ is transmitted from the dealer to the client as follows
\begin{enumerate}
	\item When a client registers with the dealer, the dealer challenges the client's intrinsic PUF with a set of inputs and stores its CRPs; \vspace{-0.05in}
	\item Before $S_k$ is distributed, the dealer selects an arbitrary CRP and uses its response $K_k$ to encode the secret with $EtM()$ to $E_k$. At the same time, the challenge $CHL_k$ takes the position of the share distribution polynomial's free coefficient. Therefore [Eq. \ref{eq:1}] becomes:
	\begin{equation} \label{eq:11}
	h_{i} = CHL_k \oplus a_{1}D_{i} \oplus a_{2}D_{i}^2 \oplus \cdots  \oplus E_kD_{i}^{t-1} .
	\end{equation}
	Then the encoded secret $E_k$ is distributed in the form of shares to the devices of the IoT system; 
	\item When $S_k$ needs to be retrieved, $t$ holders will turn in their IDs and shares to the client; \vspace{-0.05in}
	\item The client uses [Eq. \ref{eq:2}] to retrieve the encoded secret $E_k$, and by another Lagrange interpolation formula the client calculates $CHL_k$: 
	\begin{equation} \label{eq:12}
	CHL_k = \bigoplus_{i=0}^{t-1} {\frac{D_{i} \cdot h_{i}}{\prod_{j=0,j \neq i}^{t-1} {(D_{i} \oplus D_{j})}}}.
	\end{equation}
	
	\item The client takes $CHL_k$ to its PUF and regenerates the corresponding response $K_k$, which is the same key used by dealer to EtM $S_k$. This $K_k$ is used to authenticate and decrypt the retrieved encoded secret $E_k$.  \hfill\(\blacksquare\) 
\end{enumerate}
\end{algorithm}

The Odysseus system (or other IoT systems) equipped with the proposed scheme will have the workflow shown in Fig. 6. 
\begin{figure}[h] 
	\begin{center} 
		\includegraphics[width=3.5in]{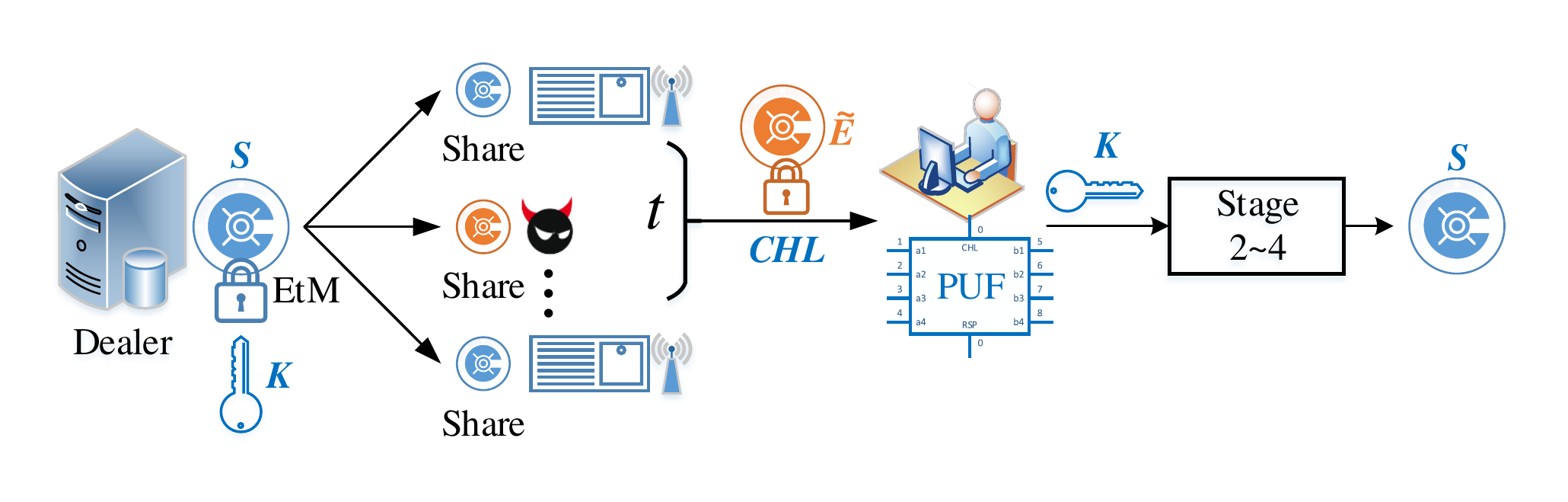} 
		\caption{\small The dealer now shares both the encoded secret and the challenge to the devices. Once the client retrieves the secret, stage 2 to 4 in Section 5.1 will be performed to identify the cheaters (if any). 
			\label{fig:proposed}} 
	\end{center} 
\end{figure} 

The advantage of this protocol is that $CHL_k$ leaks no information about $K_k$. Even if there are $t$ or more cheaters calculate $CHL_k$, they are still not able to acquire the corresponding $K_k$ because the CRPs of a PUF are not predictable. Moreover, if the PUF module generates an erroneous $\tilde{K_k}$ due to aging, temperature variations or device instability, the fuzzy extractor is able to correct the error using its ECC feature. 

When a new secret $S_l$ is about to be distributed, the dealer can select another CRP of the PUF to EtM the secret, and embed the new challenge $CHL_l$ to [Eq. \ref{eq:11}]. This makes the update of the key to $K_l$ simple and secure. 

\subsubsection{The Selection of PUFs for Secret Sharing} 
Based on where the variation comes from, there are multiple types of PUFs. Delay PUFs and memory PUFs are the two popular implementations. A delay PUF uses the random variation in delays of wires and gates, and their race condition to generate the response bits. A memory PUF is based on the random initial state (1 or 0) of each memory cell. 

Based on the size of the challenge-response pairs, there are weak and strong PUFs, which have different applications in security. Weak PUFs' CRP size grows linearly with the PUF size, while strong PUFs' CRPs grow exponentially. 

In our design, we consider the frequent updates of the key (up to one key per secret). Thus, we have selected the delay PUFs because of their large sets of CRPs. We use FPGAs to implement the secret sharing system with both the Ring Oscillator (RO) PUF based on the race condition of two ROs, and the Arbiter PUF based the delay difference between two MUX chains [\cite{MS2010}]. We also improved the design of both to increase the Hamming distance among the responses, while developing a design automation tool (introduced in Section \ref{sec:Automation}).

\subsection{Cheater Identification by Group Testing}
\label{sec:cheater-identification}
In Section 3.2.2 we pointed out that secret verification alone does not identify the cheaters nor help to retrieve the correct secret. Therefore in this paper we propose adaptive group testing, which works together with secret verification for cheater identification. It can locate up to $c_{est} = n - t$ number of cheaters, which is the theoretical upper bound. This means that in a $t$-threshold secret sharing scheme, among all the $n$ shareholders participating in our scheme, our scheme needs as few as $t$ honest parties to retrieve the correct secret. The test construction follows. 

\begin{construction} \label{cont_M}
	\normalfont For any $t$-threshold secret sharing scheme, suppose among $n$ holders there are $c$ attackers where $ 0 \leq c \leq n-t $. A test pattern to identify the honest holders and attackers can be constructed as a binary matrix $M$ of size $T \times n$, where $T$ is the number of tests needed at most. The rows of $M$ consist of all of the different $n$-bit vectors with exactly $t$ 1's and so $T = \binom{n}{t}$. Each column of the matrix therefore has $\binom{n-1}{t-1}$ number of 1's. The 1's in each row (test) correspond to the shareholders participating in that particular test. Each test is a two-step procedure: 
	\begin{enumerate}
		\item A secret reconstruction using [Eq. \ref{eq:2}] to retrieve the secret $\tilde{E}$ with its specific participants;
		\item An authentication using [Eq. \ref{eq:7} or \ref{eq:9}] over $\widetilde{E}$ to verify the validity of the retrieved secret. 
	\end{enumerate}
	The test syndrome is a $T$-bit binary vector $u$, where 0's in $u$ indicate the equality of [Eq. \ref{eq:7} or \ref{eq:9}], and 1's the inequality. \hfill\(\blacksquare\) 
\end{construction}

Then the cheater identification algorithm is:
\begin{algorithm}
	\normalfont For any $t$-threshold secret sharing scheme and its corresponding group testing matrix $M$ there are $n$ shareholders participating in the tests indexed by $H = \{0, 1, 2, \cdots, n-1 \}$ . Among the $n$ shareholders there are $c_{est}$ cheaters where $ n/3 \leq c_{est} \leq n-t $. Let $w = (w_0, w_1, \cdots, w_{n-1})$ be a $n$-digit vector and $w = u^{\top} \times M$, where $u$ is the $T$-bit binary test syndrome and $\times$ is the multiplication of regular arithmetic. The cheaters' indexes belong to the set $\{ l | \ w_l = \binom{n-1}{t-1} \}$. and the rest of the holders are honest. \hfill\(\blacksquare\)
\end{algorithm}

However, the testing technique above requires $\binom{n}{t}$ tests in total to identify the cheaters. This can be a large number when $n$ and $t$ are large. Therefore, its adaptive form, given below, drastically reduces the average number of tests to a linear formula.

\begin{algorithm} \label{alg:adaptive}
	\normalfont For a test pattern $M$ of size $T \times n$ generated by Construction \ref{cont_M}, $\triangle T$ is the number of tests needed to find the first 0 (equality of [Eq. \ref{eq:7} or \ref{eq:9}]) in the test syndrome. The $n$ shareholders are indexed by $H = \{0, 1, 2, \cdots , n-1\} $. The $t$ honest holders identified by this test are indexed by $I = \{i_0, i_1, \cdots, i_{t-1} \}$. The system only needs to run at most $n-t$ more tests whose participants are $\{i_0, i_1, \cdots, i_{t-2}, j\}$, where $j \in H \backslash I $. Each test's syndrome indicates holder $j$ as an attacker or not by 1 or 0. The total number of tests needed to identify all holders is then at most $\triangle T + (n-t)$. \hfill\(\blacksquare\)
\end{algorithm}

\subsection{Extra Invitation Module}
If the group testing module in Stage 4 cannot successfully identify the $c_{act}$ cheaters in the system, where $n-t < c_{act} \leq n$, then the number of honest shareholders is less than $t$. 

At this point, our scheme will still raise the cheating alarm based on the secret authentication. Moreover, the protocol is adaptive enough to be extended to a further stage to include an invitation module. This module can pull in the execution of the protocol additional participants and perform new rounds of group testing. From the hardware perspective, the invitation module can be power-gated and disabled when not in use. 

\begin{algorithm}
	\normalfont Let the number of honest shareholders in the current group testing be $\triangle t$ and $ 0 \leq \triangle t < t $. Suppose the system is able to identify an extra set of $t$ honest shareholders from another group. Then these $t$ honest parties can be combined into the current group with the modified group testing matrix of size $ \binom{n+t}{t} \times (n+t) $. With this new test pattern, the $\triangle t + t$ honest shareholders can be identified and the rest will be properly labeled as cheaters.\hfill\(\blacksquare\) 
\end{algorithm}

\subsection{Numeric Examples} \label{example}
Here we present two illustrative examples to demonstrate the security of the proposed protocol. The first one will be under a passive attack and the second one under an active attack. 
\begin{example}
	\normalfont For an Odysseus system equipped with the proposed secret sharing scheme, there are $t$ cheaters who want to compute the original secret $S$ stealthily. However, they can only acquire $E = EtM(K, S)$ and $CHL$. Without the client's PUF, they are not able to have the response $K$ to $CHL$. Therefore, $S$ still remains unknown to the $t$ curious cheaters. \hfill\(\square\)
\end{example}

In the second example, for simplicity we will not perform the encryption function $ENC()$ in the EtM. For the MAC function, we will use $AMD()$, since it is able to work with very short vectors. Thus, this numeric example will be relatively small and easy to follow.

\begin{example}
	\label{rasss-example}
	\normalfont We start with a share distribution among the boards of a seven-board Odysseus system configuration. We deploy on the system the proposed secure TSS scheme which is $t$-threshold and $t=3$. The original secret is a digital signature $S \in GF{(2^{12})}$ where $S = 001111110000 = 0x3F0$. The RS decoder in this scheme is constructed under the assumption that there are at most 2 cheaters. However, in the actual scenario there 4 devices which have been compromised by the cheaters.
	
	\textbf{\textit{Stage 1: Secret Encoding and Share Distribution}} \\
	The original secret $0x3F0$ is first encoded by the AMD encoding equation [Eq. \ref{eq:AMD}]. Using \text{Definition 3.2} we choose $b=4$ such that the encoding and decoding are over $GF(2^4)$, $m=1$ such that the random vector has only one symbol, and $g=3$ such that $S$ is partitioned into 3 symbols $S = (S_0, S_1, S_2)$ where $S_0 = 0x3, S_1 = 0xF$, and $S_2 = 0x0$. Suppose the dealer has chosen a response from the client's PUF which is $K = 0x0006$ whose corresponding challenge is $CHL = 0xAAAA$. The original secret will be encoded to an AMD codeword $E = AMD(K, S)$ by: 
	\begin{equation*}
	AMD(K, S) = S_0K \oplus S_1K^2 \oplus S_2K^3 = 0x1 \Rightarrow E = (0x3F01).
	\end{equation*}
	
	Then with the share distribution polynomial:
	\begin{equation*}
	h_{i} = CHL \oplus a_{1}D_{i} \oplus ED_{i}^2
	\end{equation*}
	where $a_1=0x5555$ is an arbitrarily chosen coefficient and $CHL, a_1, E \in GF(2^{16})$, this encoded secret is shared to seven Odysseus boards with IDs and shares $\{D_i : h_i\}$ = $\{1 : 0xC0FE\}$, $\{2 : 0xFC04\}$, $\{3 : 0x9650\}$, $\{4 : 0x0FB4\}$, $\{5 : 0x65E0\}$, $\{6 : 0x591A\}$, $\{7 : 0x334E\}$. 
	
	However, devices $\{3, 4, 6, 7\}$ have been compromised by cheaters and they have selected another secret $\tilde{S}=0xABCD$ and forged another share distribution polynomial: 
	\begin{equation*}
	\tilde{h}_{i} = 0xAAAA \oplus 0x7777D_{i} \oplus 0xABCD\cdot D_{i}^2.
	\end{equation*}
	
	By their IDs, their shares are changed to: $\{3 : 0x2686\}$, $\{4 : 0xDBAF\}$, $\{6 : 0x9A2F\}$, $\{7 : 0x4695\}$. 
	
	\textbf{\textit{Stage 2: Secret Reconstruction and Verification}}\\
	First, let us assume that Odysseus devices $\{2, 3, 4\}$ are selected to reconstruct the secret, of which  $\{3, 4\}$ are cheaters. By the secret reconstruction [Eq. \ref{eq:2}] the retrieved secret is
	\begin{equation*}
	\widetilde{E} = 0x5522. 
	\end{equation*}
	
	The reconstructed secret will be verified by the AMD decoder using [Eq. \ref{eq:9}]: $\widetilde{AMD(K, S)} \stackrel{?}{=}  AMD(\widetilde{K},\widetilde{S})$. Through the computation over $GF(2^4)$ we have the following inequality:
	\begin{equation*}
	\widetilde{AMD(K, S)} 
	\neq  [AMD(\widetilde{K},\widetilde{S}) = \widetilde{S_0}\widetilde{K} \oplus \widetilde{S_1}\widetilde{K}^2 \oplus \widetilde{S_2}\widetilde{K}^3].
	\end{equation*}
	Thus, cheating is detected and Stage 3 will be initiated under the assumption of $c_{est}=2$ cheaters. 
	
	\textbf{\textit{Stage 3: Share Error Correction}}\\
	Under the RS decoder, $n = 3c_{est}+1 = 7$ shareholders will be involved and up to 2 shares can be corrected using an $(n, t, d) = (7, 3, 5)$ RS code. However, there is a total number of $c_{act}=4$ cheaters $\{3, 4, 6, 7\}$, which is beyond the capability of this RS decoder. Therefore, the protocol moves to its fourth stage upon the failure of error correction.
	
	\textbf{\textit{Stage 4: Group Testing}}\\
	This stage is designed under the assumption that among all the 7 Odysseus boards from Stage 3, only $t = 3$ are not compromised by cheaters. The group testing matrix $M$ of size $T \times n$ can be constructed with \text{Construction 5.1}, where $T = \binom{n}{t} = 35, n = 7$. To save space $M$ is listed in its transposed form $M^{\top}$: 
	\newcommand\hlt[1]{\tikz[overlay, remember picture,baseline=-\the\dimexpr\fontdimen22\textfont2\relax]\node[rectangle, fill=blue!50, fill opacity=0.3, text opacity=1,] {$#1$};} 
	\[ \resizebox{\linewidth}{!}{%
		$
		\begin{array}{c|ccccc:ccccc:ccccc:ccccc:ccccc:ccccc:ccccc|ccccc}
		& 1 & 2 & 3 & 4 & 5 & 6 & 7 & 8 & 9 & 10 & 11 & 12 & 13 & 14 & 15 & 16 & 17 & 18 & 19 & 20 & 21 & 22 & 23 & 24 & 25 & 26 & 27 & 28 & 29 & 30 & 31 & 32 & 33 & 34 & 35 \\ \hline
		1 & \hlt{1} & 0 & 0 & 0 & 0 & \hlt{1} & \hlt{1} & \hlt{1} & \hlt{1} & \hlt{1} & \hlt{1} & \hlt{1} & \hlt{1} & \hlt{1} & \hlt{1} & \hlt{1} & \hlt{1} & \hlt{1} & \hlt{1} & 0 & 0 & 0 & 0 & 0 & 0 & 0 & 0 & 0 & 0 & 0 & 0 & 0 & 0 & 0 & 0\\
		
		2 & \hlt{1} & \hlt{1} & 0 & 0 & 0 & \hlt{1} & \hlt{1} & \hlt{1} & \hlt{1} & 0 & 0 & 0 & 0 & 0 & 0 & 0 & 0 & 0 & 0 & \hlt{1} & \hlt{1} & \hlt{1} & \hlt{1} & \hlt{1} & \hlt{1} & \hlt{1} & \hlt{1} & \hlt{1} & 0 & 0 & 0 & 0 & 0 & 0 & 0 \\
		
		3 & \hlt{1} & \hlt{1} & \hlt{1} & 0 & 0 & 0 & 0 & 0 & 0 & \hlt{1} & \hlt{1} & \hlt{1} & \hlt{1} & 0 & 0 & 0 & 0 & 0 & 0 & \hlt{1} & \hlt{1} & \hlt{1} & 0 & 0 & 0 & 0 & 0 & 0 & \hlt{1} & \hlt{1} & \hlt{1} & \hlt{1} & \hlt{1} & 0 & 0 \\
		
		4 & 0 & \hlt{1} & \hlt{1} & \hlt{1} & 0 & \hlt{1} & 0 & 0 & 0 & \hlt{1} & 0 & 0 & 0 & \hlt{1} & \hlt{1} & \hlt{1} & 0 & 0 & 0 & 0 & 0 & 0 & \hlt{1} & \hlt{1} & \hlt{1} & 0 & 0 & 0 & \hlt{1} & \hlt{1} & 0 & 0 & 0 & \hlt{1} & \hlt{1} \\
		
		5 & 0 & 0 & \hlt{1} & \hlt{1} & \hlt{1} & 0 & \hlt{1} & 0 & 0 & 0 & \hlt{1} & 0 & 0 & \hlt{1} & 0 & 0 & \hlt{1} & \hlt{1} & 0 & \hlt{1} & 0 & 0 & \hlt{1} & 0 & 0 & \hlt{1} & \hlt{1} & 0 & 0 & 0 & \hlt{1} & \hlt{1} & 0 & \hlt{1} & 0 \\
		
		6 & 0 & 0 & 0 & \hlt{1} & \hlt{1} & 0 & 0 & \hlt{1} & 0 & 0 & 0 & \hlt{1} & 0 & 0 & \hlt{1} & 0 & \hlt{1} & 0 & \hlt{1} & 0 & \hlt{1} & 0 & 0 & \hlt{1} & 0 & \hlt{1} & 0 & \hlt{1} & \hlt{1} & 0 & \hlt{1} & 0 & \hlt{1} & 0 & \hlt{1} \\
		
		7 & 0 & 0 & 0 & 0 & \hlt{1} & 0 & 0 & 0 & \hlt{1} & 0 & 0 & 0 & \hlt{1} & 0 & 0 & \hlt{1} & 0 & \hlt{1} & \hlt{1} & 0 & 0 & \hlt{1} & 0 & 0 & \hlt{1} & 0 & \hlt{1} & \hlt{1} & 0 & \hlt{1} & 0 & \hlt{1} & \hlt{1} & \hlt{1} & \hlt{1} \\
		\hline
		\end{array}
		$
	}\]
	
	Each test involves 3 boards and the secret retrieved by them is to be verified by [Eq. \ref{eq:9}]. Since boards $\{1, 2, 5\}$ are not compromised by cheaters, test 7 is the first test with syndrome 0. 
	
	Based on the adaptive \text{Algorithm 5.3}, $\triangle T = 7$. The system will only need to run the tests of $\{1, 6, 8, 9\}$ whose participants are boards $\{1, 2, j\}$ where $j \in H \backslash I = \{3, 4, 6, 7\}$. Thus only tests $\{8, 9\}$ are left to run. The actual number of implemented tests are then $9 < \triangle T + (n-k) \ll \binom{n}{k} = 35$. 
	
	In this way, the Odysseus boards that have been hijacked by cheaters are identified as $\{3, 4, 6, 7\}$. The functional boards $\{1,2,5\}$ will be able to retrieve the encoded legal secret $E = 0x3F01$ and, therefore, the correct digital signature $S = 0x3F0$. \hfill\(\square\)
\end{example}

\section{Design Evaluation and Automation}
\label{sec:Automation}

In this section we will evaluate the proposed scheme and offer a design automation tool for it. 

\subsection{Mis-detection Probability}
In the previous example, the AMD code works over $GF(2^4)$, where the error mis-detection probability is  $\overline{P_{miss}} = \frac{3}{2^4}$ in the worst case. To increase the security level, one can simply have the protocol work over a larger field. If the system uses HMAC as the $MAC()$ function, then $P_{miss}$ is a fixed value close to 0. Therefore, we will only test the performance of the $AMD()$ under different block sizes. 

In our experiments, $n/3 \leq c_{act} \leq n-t$. The sizes of the encoded secret $E$ are set to $\{8, 16, 32, 48, 64, 80, 96, 128\}$ bits, which are the cases for most real-world applications. Therefore, the AMD codes are over $GF(2^b)$ fields where $b \in \{2, 4, 8, 12, 16, 20, 24, 32\}$. A comparison is made between the experimental $P_{miss}$ (under $4 \cdot 2^{b}$ rounds of attack and defense) and the theoretical $\overline{P_{miss}}$.
\begin{figure}[h] 
	\begin{center} 
		\includegraphics[width=3.25in]{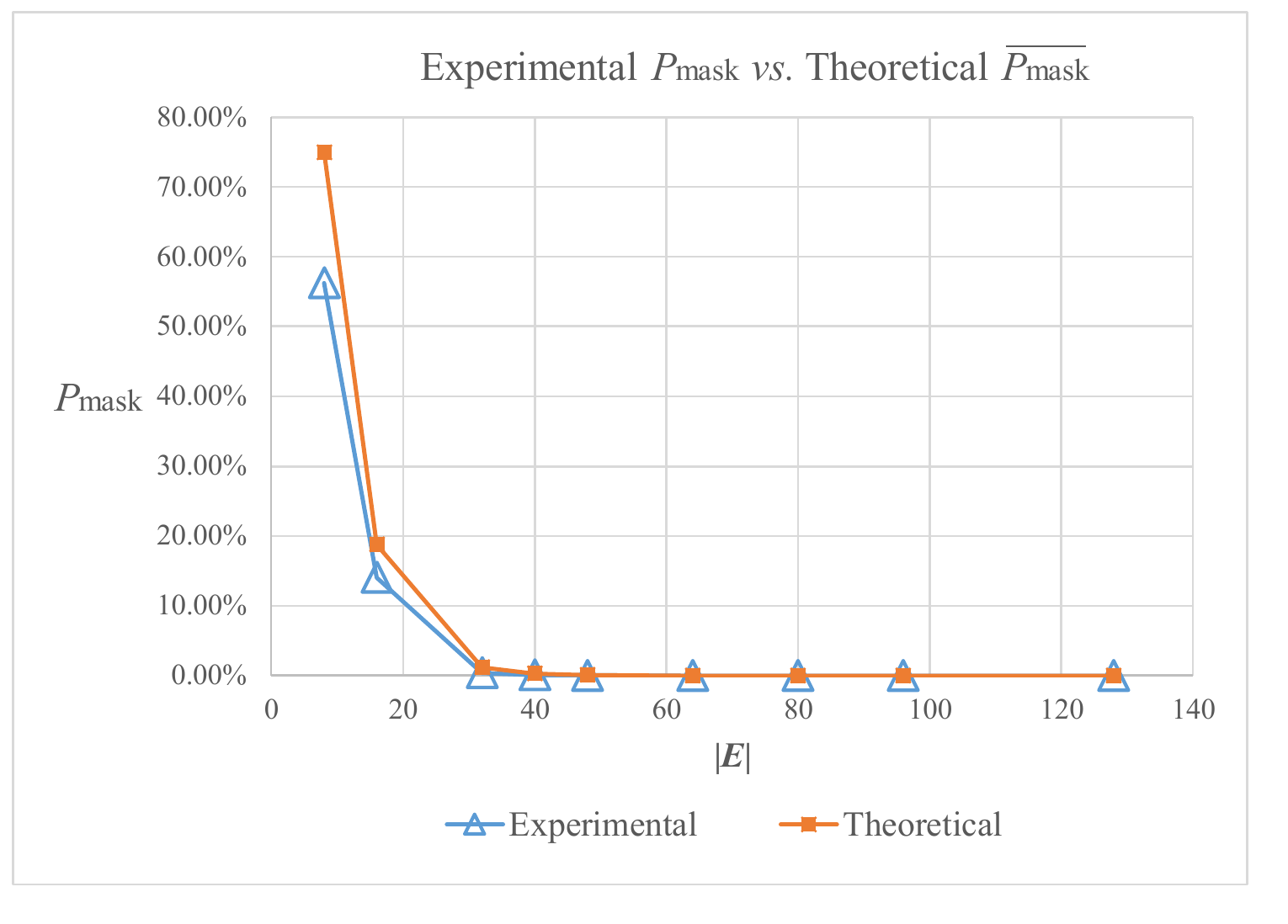} 
		\caption{\small The experimental $P_{miss}$ matches the theoretical upper bound $\overline{P_{miss}} = \frac{g}{2^b}$. The experimental results are usually better than the upper bound because [Eq. \ref{eq:9}] does not always have $h$ solutions in the finite field. Also when $b \geq 32$ the experiments did not miss a single attack. \label{fig:Exp_Theo}} 
	\end{center} 
\vspace{-0.2in}
\end{figure}

\subsection{Hardware and Runtime Overheads}
In this subsection we evaluate the complexity of the proposed scheme under $c_{act} < n/3$ and $n/3 \leq c_{act} \leq n-t$, and the hardware and runtime overheads between these two settings. The hardware cost is measured on a Xilinx Vertex 7 XC7VX330T FPGA board, and the timing on an Intel\textsuperscript{\textregistered} Core\texttrademark \ i7-6700 @ 3.4GHz and 8 GB memory machine running Linux OS. 

\begin{table}[htp]
	\resizebox{\linewidth}{!}{%
		$
		\centering
		\begin{threeparttable}
		\renewcommand{\arraystretch}{1.2}
		\caption{\large Hardware and Runtime Evaluations} \label{tab:overhead}
		\begin{tabular}{|c||c|c|c||c|c|c||}
		\hline 
		$ \boldsymbol{E}$ & \multicolumn{3}{c||}{\textbf{Hardware} (Slices)} & \multicolumn{3}{c||}{\textbf{Timing} ($10^6$ clock cycles)} \\
		(bits)  & $c_{act} < n/3$ & $c_{act} \geq n/3$ & Overhead & $c_{act} < n/3$ & $c_{act} \geq n/3$ & Overhead  \\
		\hline 
		\hline 
		{\textbf{8}} & 521  & 828  & 0.59 & 0.47 & 3.50 & 7.38   \\
		\hline 
		{\textbf{16}} & 1492  & 2256  & 0.51  & 0.56 & 5.13 & 9.17  \\
		\hline 
		{\textbf{32}} & 3977  & 6164  & 0.55 & 1.36 & 14.65 & 10.75   \\
		\hline 
		{\textbf{48}} & 6114  & 9462  &  0.55 & 1.89 & 22.34 & 11.81   \\
		\hline 
		{\textbf{64}} & 8462   & 12749  & 0.51 & 2.55 & 27.37 & 10.75  \\
		\hline 
		{\textbf{80}} & 9895  & 15804  & 0.59 & 3.18 & 32.47 & 10.21  \\
		\hline 
		{\textbf{96}} & 11873  & 18918  & 0.59  & 3.68 & 40.90 & 11.12   \\
		\hline 
		{\textbf{128}} & 17842 & 27695  & 0.55 & 4.79 & 50.05 & 10.44  \\
		\hline 
		\end{tabular}
		\begin{tablenotes}
		\begin{footnotesize}
		\small
		\item  [I] With only 60\% of the hardware overhead, the the cheater tolerance capability can be drastically improved. 
		\item  [II] The timing overhead is efficiently reduced by Algorithm \ref{alg:adaptive}.
		\end{footnotesize}
		\end{tablenotes}
		\end{threeparttable}
		\vspace{-0.1in}
		$
	}
\end{table}

\subsection{Design Automation}
Although one can manually make a secret sharing system with PUF on FPGAs, it still involves a good amount of work between writing the HDL code, fixing the routing and placement of PUF's basic elements, configuring the bitstream, and so forth. Also, with a change in one parameter, the entire system may need to be modified. Therefore, we have designed an automation tool that simply takes the user's inputs of four parameters (secret size, security level (for AMD only, HMAC default as $2^{-128}$), total number of holders $n$, threshold $t$, and MAC function). In addition, we also provide a PUF automation tool to generate the PUFs based on user specified response and challenge sizes. 

In this tool, the system's HDL codes and PUF's fixed-routing configuration are pre-written in a folder named ``Templates." The tool will generate the system according to user specified parameters based on the files in this folder. For any future modification of the system, only the templates need to be adjusted, and the generator tool can stay unchanged.  
\begin{figure}[h] 
	\begin{center} 
		\includegraphics[width=3.5in]{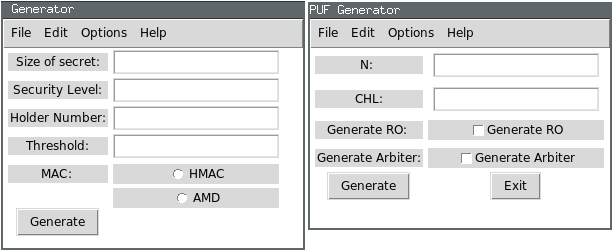} 
		\caption{\small The GUIs for the secret sharing system generator (left) and the PUF generator (right). With this tool any research can generator his own customized secret sharing system in a few clicks to assist his/her researches on secret sharing and PUF.  
			\label{fig:GUI1}} 
		\vspace{-0.4in}
	\end{center} 
\end{figure}

\section{Conclusion}
In this paper we have proposed a secure and robust scheme to share confidential information in IoT systems. This scheme uses Threshold Secret Sharing (TSS) to split the information into shares to be kept by all devices in the system, so that the malfunction of a single device will not harm the security of the entire system. In case of more erroneous or rogue devices, this scheme ensures both the privacy and integrity of that piece of information even when there is a large number of sophisticated and coordinated attackers hijacking the devices. The scheme is able to identify all of the compromised devices, while still keeping the secret unknown to, and unforgeable by, the attackers. In contrast, earlier secure schemes suffer from the leakage of secrets, the forgery of fake secrets, or even the misidentification of the honest devices as cheaters. This scheme works in an adaptive manner, such that a more powerful (and power-consuming) security module will only be activated when the previous modules fail. Therefore, the average power consumption is minimized. This scheme also applies to other IoTs with a structure similar to the Odysseus.

\begin{wrapfigure}{l}{20mm} 
	\includegraphics[width=1in, height=1in, clip, keepaspectratio]{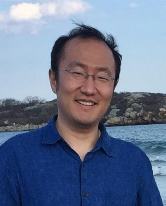}
\end{wrapfigure} 
\textbf{Lake Bu} is a PhD candidate in the Adaptive and Secure Computing Lab at Boston University. His research focus is on hardware reliability and security. His goal is to design reliable, secure, and robust hardware systems for distributed systems, whose communications' confidentiality and authenticity can be ensured. He studies various types of threats on hardware systems such as random errors, error injection attacks, Man-In-The-Middle attacks, and collusive attacks. He explores defense approaches using security oriented codes, secure secret sharing schemes, and cryptography etc. on FPGA platforms.  

\bigskip 

\begin{wrapfigure}{l}{20mm} 
	\includegraphics[width=1in,height=1in,clip,keepaspectratio]{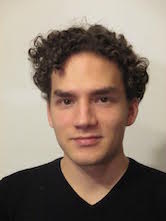}
\end{wrapfigure} 
\textbf{Mihailo Isakov} is a PhD candidate in the Adaptive and Secure Computing Lab at Boston University. His research focus is on investigating neuromorphic computing systems. In particular, he explores architectural support for machine learning algorithms prone to parallelization and efficient hardware implementation. Currently, he is investigating novel neural architectures and accelerating deep neural networks by constraining them to fast, low-power and highly parallelizable operations.   

\bigskip

\begin{wrapfigure}{l}{20mm} 
	\includegraphics[width=1in,height=1in,clip,keepaspectratio]{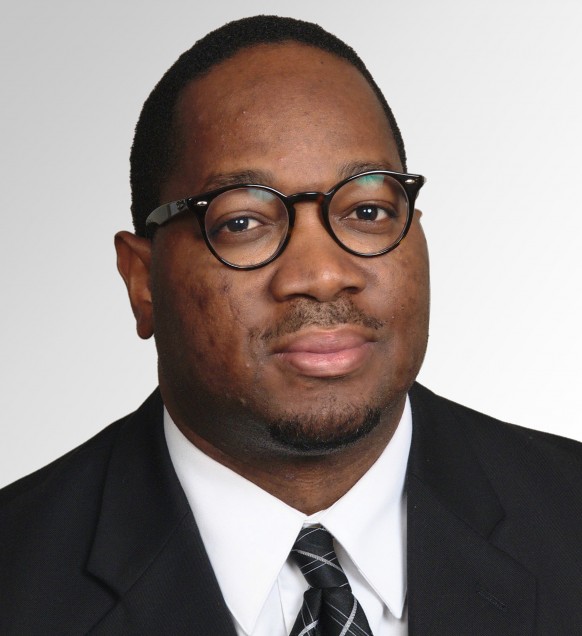}
\end{wrapfigure}\par
\textbf{Michel A. Kinsy} is an Assistant Professor in the Department of Electrical and Computer Engineering at Boston University (BU), where he directs the Adaptive and Secure Computing Systems (ASCS) Laboratory. He focuses his research on computer architecture, hardware-level security, neural network accelerator designs, and cyber-physical systems. Dr. Kinsy is an MIT Presidential Fellow, the 2018 MWSCAS Myril B. Reed Best Paper Award Recipient, DFT'17 Best Paper Award Finalist, and FPL'11 Tools and Open-Source Community Service Award Recipient. He earned his PhD in Electrical Engineering and Computer Science in 2013 from the Massachusetts Institute of Technology. \par \smallskip

\end{document}